# Teoria Gravitazionale e Applicazioni Cosmologiche: Possibili Sviluppi Futuri

Maurizio Gasperini

Dipartimento Interateneo di Fisica , Università di Bari
Istituto Nazionale di Fisica Nucleare, Sezione di Bari, Bari, Italy

Una rapida rassegna di alcuni tra i principali argomenti che riguardano l'interazione gravitazionale e la cosmologia, e che sono attualmente in corso di studio e di dibattito. Si parlerà in particolare del principio di equivalenza, di materia oscura, di energia oscura, della costante cosmologica, di gravità quantistica, e dell'Universo primordiale.

## 1. Piccole e grandi distanze

Come in tutti i campi della fisica (teorica e sperimentale), anche nell'ambito della gravitazione e della cosmologia ci sono diversi problemi attualmente aperti, alcuni dei quali decisamente importanti. Essendo di professione un fisico teorico, in questa breve nota mi concentrerò sulla discussione di tali problemi e sulla loro possibile evoluzione futura in un contesto puramente teorico/fenomenologico.

Possiamo subito fare una prima - e apparentemente ovvia - distinzione tra gli effetti gravitazionali da scoprire (o semplicemente da verificare) alle piccole distanze e quelli alle grandi distanze (come vedremo in seguito, però, tale separazione è valida se le distanze non sono troppo piccole o troppo grandi).

Sappiamo infatti che alle ordinarie distanze macroscopiche la gravità è molto ben descritta dalla relatività generale di Einstein, che si riduce alla teoria di Newton nel limite in cui tutti gli effetti relativistici sono trascurabili. La teoria di Einstein è basata sul principio di equivalenza, ossia, da un punto di vista più formale, sull'ipotesi che l'interazione gravitazionale sia completamente descritta dallo scambio di particelle di spin 2 e massa nulla, i gravitoni (si veda ad esempio [1]), che si accoppiano in modo universale - ossia, con la stessa intensità - a tutte le forme di energia. Ma cosa succede a distanze molto pic-cole o molto grandi, non direttamente accessibili agli attuali *test* del principio di equivalenza [2]?

A piccole distanze potrebbe entrare in gioco una ulteriore componente dell'interazione gravitazionale mediata da nuove particelle massive, che producono forze a corto raggio d'azione (e sono quindi invisibili a livello macroscopico) e che dipendono dalla struttura interna dei corpi (ad esempio, dalla loro carica barionica). A grandi distanze, invece, potrebbe esserci il contributo di altre particelle, con massa nulla o talmente piccola da generare forze a lunghissimo raggio, che si accoppiano diversamente ai vari tipi di materia ma così debolmente da produrre effetti sensibili solo quando l'ordinaria gravità macroscopica è trascurabile (ad esempio, a livello cosmico).

In entrambi i casi la teoria gravitazionale di Einstein andrebbe modificata, e in modo diverso a seconda delle distanze coinvolte. Per quel che riguarda il prossimo futuro (diciamo i prossimi dieci anni) personalmente non mi aspetto che vengano trovate violazioni del principio di equivalenza (o delle equazioni gravitazionali standard) medianti esperimenti effettuati a distanze minori di quelle che riguardano le attuali verifiche, che si spingono fino ad una scala di circa $10^{-3}$ cm [3]. Se andiamo a distanze molto più piccole entriamo infatti nel regime molecolare, atomico, nucleare, etc, dove le interazioni dominanti sono svariati ordini di grandezza più intense di quella gravitazionale, e gli effetti della gravità diventano fisicamente invisibili.

## 2. Gravità cosmica oscura

Per quel che riguarda la gravità a grandi distanze, invece, sono più possibilista. In quel caso, infatti, non si tratta (solo) di effettuare verifiche sperimentali del principio di equivalenza, ma anche di confrontare le previsioni della teoria standard con la dinamica delle sorgenti gravitazionali dominanti su grandi scale. E se ci spingiamo alle massime scale cosmiche osservabili (scala galattica, intergalattica e scala di Hubble) abbiamo vari interrogativi attualmente aperti, tra cui, in particolare, quelli che riguardano la corretta interpretazione delle quantità denominate `materia oscura` e `energia oscura`.

Tutti i tentativi finora effettuati di identificare queste sorgenti di gravità cosmica come composte di particelle, o come campi, previsti dal modello standard (o dalle sue generalizzazioni) sono da una parte certamente compatibili con le attuali osservazioni, ma, dall'altra parte, non si è ancora trovata una indiscutibile prova (diretta o indiretta) che sia a favore di uno specifico modello e smentisca tutte le altre possibili interpretazioni.

Per cui, va anche seriamente considerata la possibilità (studiata da molti gruppi di ricerca) che le modifiche del modello cosmologico standard, prodotte da esotiche componenti `oscure` introdotte per mantenere invariate le equazioni gravitazionali a livello cosmico, siano invece prodotte da una dinamica gravitazionale cosmica che è diversa da quella macroscopica, e che non può essere correttamente (o completamente) descritta dalle equazioni della teoria di Einstein.

Per riportare solo qualcuno dei possibili e numerosi esempi, la corretta teoria che non richiede la presenza di materia oscura potrebbe essere basata su un modello di spazio-tempo che a grandi scale ha una struttura geometrica di tipo frattale anzichè Riemanniana [4]; oppure, a grandi scale, potrebbero entrare in gioco nuovi e diversi contributi della geometria all'interazione gravitazionale con la materia [5]; e così via. Io credo, o perlomeno auspico, che interrogativi di questo tipo possano venire almeno parzialmente risolti grazie alla precisione sempre crescente delle osservazioni astronomiche che sono in corso, e che sono previste per il prossimo futuro.

## 3. Costante cosmologica

In questo contesto si ritrova anche quello che probabilmente è uno dei problemi più famosi non solo della teoria gravitazionale ma di tutta l'attuale scienza fisica: il problema della costante cosmologica. La componente cosmica di energia oscura, infatti, è stata inventata proprio per evitare di inserire direttamente una costante cosmologica nelle equazioni di Einstein. Ma perché dovremmo evitare questa semplice operazione formale?

Perché la costante cosmologica rappresenta, fisicamente, la densità d'energia totale del vuoto. Se cerchiamo di calcolarla, includendo il contributo di tutte le interazioni descritte dalle attuali teorie unificate, troviamo un valore esattamente nullo se il vuoto gode di importanti simmetrie come la `supersimmetria` tra campi bosonici e fermionici. Se tali simmetrie sono rotte, invece, si ottiene per la costante cosmologica un valore talmente elevato da risultare inaccettabile (per ottenere il valore giusto bisognerebbe far ricorso a modelli di spazio-tempo molto esotici [6]). Ecco dunque un tipico esempio di effetto che impedisce (come preannunciato) una netta distinzione tra la fisica delle piccole e delle grandi distanze: infatti, l'energia di punto zero dei campi quantistici a livello microscopico, che contribuisce all'energia del vuoto, potrebbe avere effetti sull'evoluzione dell'intero Universo a livello cosmico.

## 4. Gravità quantistica

Tornando alle piccole distanze, non posso non menzionare quello che a mio avviso è il più importante problema aperto: la formulazione di una corretta teoria gravitazionale quantistica. Per affrontare (ed eventualmente risolvere) tale problema è necessario riferirsi a scale di distanze non solo molto più piccole di quelle che entrano in gioco nella verifica del principio di equivalenza, ma anche molto più piccole delle minime distanze accessibili con gli attuali acceleratori di altissima energia (circa $10^{-16}$ cm).

Una consistente e completa quantizzazione della gravità, infatti, è inevitabilmente e strettamente legata alla soluzione di altri enigmi che riguardano ad esempio la presenza di nuove (super-)simmetrie, di dimensioni spaziali e/o temporali extra (quante?), di una struttura discreta della geometria microscopica, di una corrispondente scala di lunghezza minima che impedisce (anche concettualmente) la nozione di oggetti puntiformi, di una possibile modifica del principio di indeterminazione. E così via.

Personalmente non mi aspetto che indizi di questa nuova fisica possano apparire alle energie (e quindi alle scale di distanza) raggiungibili dagli acceleratori di particelle disponibili ora e in un prossimo futuro. La ragione, a mio avviso, è che non può esistere un modello di gravità quantistica esatto, consistente a tutti gli ordini di approssimazione, e nel quale gli effetti quantistici della gravità diventino confrontabili con quelli delle altre interazioni, se non nell'ambito di uno schema teorico che descriva in modo unificato tutte le interazioni fondamentali.

L'unico schema teorico attualmente in grado di realizzare questa unificazione - perlomeno in linea di principio - è la teoria delle `(super)stringhe`. E così come l'unificazione delle interazioni elettromagnetiche, deboli e forti è caratterizzata dalla cosiddetta scala di `grande unificazione`, ossia da scale di energia dell'or- dine dei $10^{15}$ GeV, l'unificazione completa che include anche la gravità è caratterizzata dalla sca- la di stringa, che si pensa corrispondere a ener- gie pari a circa un decimo di quella di Planck, ossia a energie dell'ordine dei $10^{18}$ GeV. Alla scala di Planck, tra parentesi, le equazioni della relatività generale diventano quantisticamente inconsistenti anche secondo l'ordinaria teoria quantistica dei campi.

Sono dunque queste, secondo l'attuale schema teorico, le scale di energia, e le corrispondenti scale di distanza (dell'ordine dei $10^{-32}$ cm), alle quali dovremo arrivare per verificare le conseguenze di un modello esatto di gravità quantistica (e distinguerlo dalle sue approssimate versioni semiclassiche, che a quelle scale non sono più valide).

## 5. Universo primordiale

E qui troviamo un altro stretto collegamento tra le piccolissime e le grandissime distanze. Se non possiamo trovare tracce di effetti gravitazionali quantistici alle piccole scale con esperimenti di laboratorio (perchè è impensabile che gli attuali rivelatori possano migliorare di tanti ordini di grandezza il loro massimo regime energetico, pari a circa $10^3$ GeV), potremmo però trovarle alle grandi scale di distanza.

Guardare lontano nello spazio, infatti, significa guardare indietro nel tempo (perchè i segnali viaggiano a velocità finita, e più lontane sono le sorgenti, più vecchie sono le informazioni che oggi riceviamo). Il nostro Universo, d'altra parte, si sta espandendo, per cui segnali più vecchi corrispondono a epoche più remote nelle quali l'Universo era molto più piccolo, e quindi più denso, più caldo e più curvo. In altri termini, i segnali più lontani ci portano fotografie di uno stato cosmico caratterizzato da una scala di ener-gia molto più elevata, e una scala di distanza molto più piccola, di quelle attuali.

Ci sono dei limiti, però, alla possibilità di raccogliere informazioni dirette sullo stato primordiale dell'Universo. Se ci basiamo, ad esempio, sui segnali elettromagnetici attualmente accessibili alla nostra informazione possiamo allora risalire al massimo fino all'epoca del cosiddetto *decoupling* (disaccoppiamento), alla quale l'Universo è diventato trasparente alla radiazione elettromagnetica, che ha potuto quindi propagarsi fino ai giorni nostri.

In tempi precedenti l'Universo era così denso che la radiazione elettromagnetica emessa veniva immediatamente riassorbita dalla materia circostante, per cui tale radiazione non è più accessibile alla nostra diretta osservazione. La scala di energia del *decoupling* è milioni di volte superiore alla attuale scala di energia cosmica, ma è purtroppo ancora molto lontana da quelle scale - menzionate in precedenza - alle quali ci aspettiamo che la gravità venga modificata dagli effetti quantistici.

Se crediamo nel modello cosmologico standard, però, l'evoluzione cosmica non parte dal *decoupling*. Possiamo andare più indietro nel tempo, trovando che l'energia, la densità, la temperatura continuano a crescere senza limiti, raggiungendo inevitabilmente (perlomeno in linea di principio) anche la scala di stringa e la scala di Planck, alle quali gli effetti gravitazionali quantistici dovrebbero essere rilevanti.

Diventa quindi cruciale chiedersi: che informazioni potremmo avere su queste epoche così remote, e precedenti al *decoupling*? In particolare, sulle fasi primordiali che l'Universo ha attraversato subito dopo (o subito prima) l'enigmatica epoca del *Big Bang*? Sottolineo per chiarezza che il termine *Big Bang* va qui inteso (come più volte ribadito anche dal collega e amico Gabriele Veneziano) non tanto in senso formale come un'ipotetica singolarità iniziale, quanto invece in senso fisico come l'epoca del cosiddetto *reheating* (riscaldamento), caratterizzata da un'esplosiva produzione di particelle di altissima energia che rapidamente si diffondono riempendo tutto lo spazio al momento disponibile, e segnando così l'inizio della fase cosmologica standard dominata dalla radiazione.

# 6. Radiazione gravitazionale fossile

Concludiamo questa breve discussione cercando di rispondere alla domanda formulata alla fine della sezione precedente. La risposta (forse sorprendentemente) è positiva: non possiamo ricevere segnali elettromagnetici da quelle epoche così remote, ma possiamo ricevere le loro fotografie stampate sulla radiazione gravitazionale che è stata emessa grazie agli effetti quantistici durante quelle epoche e che, al contrario della radiazione elettromagnetica, si è potuta propagare liberamente fino ai giorni nostri. Tale radiazione potrebbe essere direttamente rivelata dalle antenne gravitazionali attualmente in funzione, o da quelle di prossima costruzione, non appena verrà raggiunta la sensibilità necessaria [7].

E in che modo questi segnali potrebbero darci informazioni sulla dinamica primordiale e sul corretto modello di gravità quantistica da usare? In un modo molto semplice. Lo spettro (ossia, l'andamento in frequenza) di questa radiazione gravitazionale fossile riproduce esattamente (come una fotografia, appunto) l'andamento nel tempo della curvatura dell'Universo durante le epoche in cui è stata prodotta.

D'altra parte, la curvatura cosmica prima del *reheating* è prevista essere decrescente nel tempo (o al massimo costante) secondo i modelli di evoluzione primordiale (e inflazionaria), basati sulla quantizzazione della Relatività Generale; crescente nel tempo, invece, secondo i modelli basati sulla teoria delle stringhe. Di conseguenza, ci possiamo aspettare un fondo di radiazione gravitazionale fossile con spettro decrescente (o piatto) in frequenza per i modelli cosmologici standard, crescente in frequenza per i modelli di cosmologia di stringa [8].

Tale differenza è cruciale perchè con uno spettro crescente l'intensità del segnale gravitazionale cresce con l'aumentare della frequenza, e quindi ha più probabilità di essere rivelato nel *range* di frequenze (relativamente alte rispetto alle tipiche frequenze cosmiche) nel quale sono sensibili le attuali antenne. Il contrario avviene invece per lo spettro del modello standard, che prevede dunque in quel *range* un segnale troppo debole per essere attualmente rivelato.

La scoperta e la misura di un fondo cosmico di onde gravitazionali fossili, residuo di epoche primordiali, potrebbe quindi darci conferma degli effetti fisici prodotti dalla gravità quantistica, nonchè importanti indicazioni per la quantizzazione e l'unificazione di tutte le interazioni fondamentali. Speriamo che ciò avvenga in un futuro non troppo lontano, in modo tale che tutti i lettori del presente articolo possano assistere e partecipare a questi importanti progressi scientifici.

## Ringraziamenti

**Maurizio Gasperini:** attualmente in pensione, è stato professore ordinario di Fisica Teorica all'Università di Bari. Ha svolto (e svolge) attività scientifica e didattica nel campo della relatività, della cosmologia, della gravitazione e della teoria delle interazioni fondamentali. È stato in precedenza ricercatore all'Università di Torino, *Academic Staff Member* alla Università della California (Santa Barbara) e *Scientific Associate* al CERN dove, in collaborazione con Gabriele Veneziano, ha formulato e sviluppato un modello di universo primordiale basato sulla teoria delle stringhe. Per ulteriori dettagli si veda `http://www.ba.infn.it/~gasperin/academic.html`

[1] M. Gasperini: *Le onde gravitazionali nella fisica moderna*, Ithaca: Viaggio nella Scienza, XII (5) 2018. `http://ithaca.unisalento.it/`

[2] M. Gasperini: *Gravity at finite temperature, equivalence principle, and local Lorentz invariance*, in "Breakdown of the Einstein's Equivalence Principle", ed. by A. G. Lebed (World Scientific, 2023, Cap. 3, p. 77).

[3] P. Touboul et al: *Space test of the equivalence principle: first results of the MICROSCOPE mission*, Class. Quantum Grav., 36 (225006) 2019.

[4] L. Cosmai, G. Fanizza, F. Sylos Labini, L. Pietronero and L. Tedesco: *Fractal universe and cosmic acceleration in a Lemaitre-Tolman-Bondi scenario*, Class. Quantum Grav., 36 (2019) 4.045007

[5] G. Fanizza, G. Franchini, M. Gasperini and L. Tedesco: *Comparing the luminosity distance for gravitational waves and electromagnetic signals in a simple model of quadratic gravity*, Gen. Rel. Grav., 52 (2020) 11.111

[6] M. Gasperini: *Higher-dimensional perturbations of the vacuum energy density*, JHEP, 06 (2008) 009.

[7] Y. Jiang, X. L. Fan and Q. G. Huang: *Search for stochastic gravitational-wave background from string cosmology with Advanced LIGO and Virgo's O1-O3 data*, preprint, arXiv:2302.03846 ([gr-qc]).

[8] M. Gasperini: *Elementary introduction to pre - big bang cosmology and to the relic graviton background*, in "Gravitational Waves", edited by I. Ciufolini, V. Gorini, U. Moschella and P. Fre' (IOP Publishing, Bristol, 2001) p. 280-337. ISBN: 0-7503-0741-2; hep-th/9907067